\documentclass[%
aip,
cp,  % Conference Proceedings
amsmath,amssymb,%nobibnotes,
reprint,%
]{revtex4-2}

\usepackage{graphicx}% Include figure files
\usepackage{dcolumn}% Align table columns on decimal point
\usepackage{bm}% bold math
\usepackage[utf8]{inputenc}
\usepackage[T1]{fontenc}
\usepackage{mathptmx,hyperref} 

\begin{document}
	
	\title{Quantum process tomography of adiabatic and superadiabatic stimulated Raman passage}
	\author{Shruti Dogra} % Write as First name Surname
%	\email[Corresponding author: ]{first.author@insitution.edu}
	\author{Gheorghe Sorin Paraoanu}%
%	\email{second.author@institution.edu.}
	\affiliation{}
	\date{\today} % It is always \today, today, but any date may be explicitly specified
	% Not printed for conference proceedings
	
	\begin{abstract}
Quantum control methods for three-level systems have become recently an important direction of research in quantum information science and technology. Here we present numerical simulations using realistic experimental parameters for quantum process tomography in STIRAP (stimulated Raman adiabatic passage) and saSTIRAP (superadiabatic STIRAP). Specifically, we identify a suitable basis in the operator space as the identity operator together with the 8 Gell-Mann operators, and we calculate the corresponding process matrices, which have $9\times 9=81$ elements. We discuss these results for the ideal decoherence-free case, as well as for the experimentally-relevant case with decoherence included. 
	\end{abstract}
	
	\maketitle

%\section{\label{sec:level1}First-level heading:\protect\\ The line
%	break was forced \lowercase{via} \textbackslash\textbackslash}

\section{Introduction  \label{sec:level1}}
\par

With rapidly advancing experimental approach to quantum computation 
and quantum information \cite{nielsen-book-2002}, and especially with newly emerging era 
of cloud-based experimental quantum computing platforms \cite{Gambetta2019,PhysRevA.94.032329}, the running of algorithms on few-qubits processors has become widely available \cite{Perelshtein2020,Maniscalco2020,Gambetta2017, PhysRevLett.122.080504,PhysRevA.95.032131,PhysRevA.94.012314,Paraoanu2018,Panigrahi2019,shukla-pla-2020}. At the core of these developments are key advancements in improving the fidelity of quantum gates, which require their precise characterization by quantum tomography.
Quantum process tomography (QPT) is a method to characterize an arbitrary quantum operation
by operating that process on a set of known test states forming a 
complete basis~\cite{chuang-jmod-1997}. Despite being
quite demanding in terms of resources, QPT is essential in quantum 
computing for specifying the accuracy of implementation of
quantum gates. There have been several theoretical proposals as well 
as experimental demonstrations to obtain quantum process 
tomography~\cite{Howard-njp-2006, martinis-nature-2010,palmieri-npj-2020}
and also to reduce the resources~\cite{shukla-pra-2014,akshay-pra-2018,PhysRevLett.100.190403,govia-nature-2020}. 
So far, most of the implementations aim at characterizing
quantum processes in multi-qubit systems. Here we work out process tomography for a three-level system (qutrit).
Multilevel processes provide an efficient way to quantum computing tasks \cite{Lanyon2009} and they are directly relevant for quantum simulations of many-body systems \cite{Paraoanu2014,Nori_review_2011}. Other possible applications in  the field of circuit QED concern operations via dissipative microwave links \cite{Cleland2020},  multilevel metrology \cite{shlyakhov_2018,danilin_2018}, the characterization of qutrit-based quantum computing architectures \cite{Siddiqi2020}, and the study of the ultrastrong coupling \cite{Falci_ultra}.

Here we show how to implement quantum process tomography for a three-level system (qutrit) undergoing adiabatic and superadiabatic processes. Adiabatic processes are an important component of the control toolbox available in quantum physics. In three-level systems, a paradigmatic transfer process is STIRAP (stimulated Raman adiabatic passage) \citep{Bergmann_2019}, which has been demonstrated with a superconducting circuit \cite{kumar-nature-2016}. The superadiabatic STIRAP (saSTIRAP) is a process whereby the standard STIRAP is augmented by a counterdiabatic Hamiltonian, which has the role of keeping the system in the dark state and which is obtained by reverse Hamiltonian engineering \cite{Torrontegui13}. This process has also been demonstrated experimentally \cite{antti-science-2019}. Although developed as a way to transfer population, STIRAP and saSTIRAP can also assist in the realization of certain quantum gates \cite{antti-NOTgate-2018} and can be applied to superpositions \cite{antti-photonics-2016}. Recently a geometrical picture of these processes has been developed, based on the Majorana star representation \citep{Dogra2020}.

%%%%%%%%%%%%%%
\section{Quantum process tomography for qutrits  \label{sec:level2}}

\paragraph{Introduction to quantum process tomography}

Let us consider a $\mathrm{d}-$dimensional quantum system undergoing a process $\epsilon$ which takes an arbitrary state: $\rho \rightarrow \epsilon(\rho)$.
This, using the operator sum representation \cite{Kraus-aop-1971, nielsen-book-2002}, may be written as
\begin{eqnarray} \label{eq:kraus}
\epsilon(\rho) = \sum_{i} E_i \rho E_i^{\dagger},
\end{eqnarray}
such that $E_i=\sum e_{im} \tilde{E}_m$, where
$\tilde{E}_m$ form a complete basis in $d^2$-dimensional operator space
and $e_{im}$ are the complex coefficients. We have
\begin{eqnarray} \label{eq:def1}
\epsilon(\rho) = \sum_{m,n} \tilde{E}_m \rho \tilde{E}_n^{\dagger} \chi_{mn},
\end{eqnarray}
where $\chi_{mn}=\sum_i e_{im}e_{in}^{\ast}$. $\chi_{mn}$ is a complex
positive Hermitian matrix that characterizes the process $\epsilon$ for
a fixed set of basis operators: $\{ \tilde{E}_m \}$. The aim is to
 obtain the process matrix $\chi$, but the complex coefficients $e_{im}$s
 are not directly accessible.
In practical situations however, the more accessible entity is
the state of the system,
which can be obtained using standard procedures of quantum state tomography
(QST)~\cite{nielsen-book-2002,suter-book-2004}. In order to characterize this process via QST-based alternative approach, the process $\epsilon$ 
is made to operate on $\mathrm{d}^2$ different 
linearly independent matrices ($\rho_i$, $i\in [ 1, \mathrm{d}^2 ]$), 
which form a complete basis to represent a quantum state $\rho$. The resultant states are obtained as
\begin{equation} \label{eq:def2}
\epsilon(\rho_j) = \sum_k \lambda_{jk} \rho_k ,
\end{equation}
where the coefficients $\lambda_{jk}$ can be obtained from the QST of 
respective states $\epsilon(\rho_j)$. Exploiting the equivalence of the 
two approaches in Eqns.~\ref{eq:def1} and~\ref{eq:def2} for 
a state $\rho_j$, we obtain
\begin{eqnarray}
\epsilon(\rho_j)=
\sum_{m,n} \tilde{E}_m \rho_j \tilde{E}_n^{\dagger} \chi_{mn},
&=& \sum_k \lambda_{jk} \rho_k \\
{\rm or} \qquad \qquad
\sum_k \sum_{m,n} \beta_{jk}^{mn} \chi_{mn} \rho_k
&=& \sum_k \lambda_{jk} \rho_k ,
\end{eqnarray}
where $\tilde{E}_m \rho_j \tilde{E}_n^{\dagger}=\sum_k \beta_{jk}^{mn} \rho_k$.
Linear independence of $\rho_k$s imply that the respective coefficients are 
to be equated, such that
\begin{equation}\label{eq:qpt}
\beta \chi = \lambda \quad  \implies \quad \chi = \beta^{-1} \lambda,
\end{equation}
where $\chi$ and $\lambda$ are $\mathrm{d}^4-$ dimensional vectors,
and $\beta$ is $d^4 \times \mathrm{d}^4-$ dimensional invertible matrix with rows 
labelled by $jk$ and columns by $mn$~\cite{nielsen-book-2002}.
The $d^4 \times 1$ dimensional $\chi$ thus obtained is reshuffled 
to obtain the  $\mathrm{d}^2 \times \mathrm{d}^2$ dimensional process matrix, which 
uniquely represents the desired process $\epsilon$ for a fixed basis.

\paragraph{The case of qutrits}

We perform the quantum process tomography for the case of a qutrit,
which is a three-level quantum system ($\mathrm{d}=3$). A set of nine matrices,
which form a complete basis to represent an arbitrary quantum state in $3\times3-$dimensional operator space is
 $|p \rangle \langle q |$, where $p,q=0,1,2$, which consist of three
 Hermitian and six non-Hermitian matrices. The procedure
 for QPT as described earlier is also valid for the non-Hermitian  
 basis, as being used here. For the real situations such 
 as experiments, these non-Hermitian matrices can be realized as a linear 
 combination of Hermitian density operators. 
 The basis set to represent the $\tilde{E}_m$'s is chosen to be 
 a $3\times3$ identity operator ($\tilde{E}_1=\mathbb{I}$) and
 the eight Gell Mann matrices ($\tilde{E}_{i+1}=\Lambda_i$, $i \in [1,8]$).  
The Gell-Mann lambda matrices are a set of 8  matrices that form a representation of the generators of 
the Lie algebra associated with the group SU(3). They are defined as 

%%%%%%%%%%%%%%%%
\begin{eqnarray}
\Lambda_1 &=& \left(\begin{array}{lll}
0 & 1 & 0 \\ 1 & 0 & 0 \\ 0 & 0 & 0
\end{array}\right), \, 
\Lambda_3 = \left(\begin{array}{lll}
1 & 0 & 0 \\ 0 & -1 & 0 \\ 0 & 0 & 0
\end{array}\right), \, 
\Lambda_5=\left(\begin{array}{lll}
0 & 0 & -i \\ 0 & 0 & 0 \\ i & 0 & 0
\end{array}\right), \,
\Lambda_7=\left(\begin{array}{lll}
0 & 0 & 0 \\ 0 & 0 & -i \\ 0 & i & 0
\end{array}\right), \nonumber \\  
\Lambda_2 &=& \left(\begin{array}{lll}
0 & -i & 0 \\ i & 0 & 0 \\ 0 & 0 & 0
\end{array}\right), \,
\Lambda_4=\left(\begin{array}{lll}
0 & 0 & 1 \\ 0 & 0 & 0 \\ 1 & 0 & 0
\end{array}\right), \, 
\Lambda_6=\left(\begin{array}{lll}
0 & 0 & 0 \\ 0 & 0 & 1 \\ 0 & 1 & 0
\end{array}\right), \,
\Lambda_8=\frac{1}{\sqrt{3}}\left(\begin{array}{lll}
1 & 0 & 0 \\ 0 & 1 & 0 \\ 0 & 0 & -2
\end{array}\right).          
\end{eqnarray}
Together with the identity matrix $\mathbb{I}_{3\times3}$, these matrices can be used to generate any unitary transformations of a qutrit. 
In order to characterize an arbitrary process $\epsilon$ in a qutrit, 
we simulate 
the qutrit in nine different initial states $|p \rangle \langle q |$.
Corresponding to each of these states states, the qutrit is then allowed
 to undergo the same process $\epsilon$. At the end of which, respective
 final states are obtained via quantum state tomography. These nine final
 states contibute $81$ complex elements, which form $81\times 1-$dimensional
 vector $\lambda$. The $81 \times 81-$dimensional $\beta$ results from
 the combinations of $\tilde{E_m}$s, finally giving rise to the process
 matrix $\chi$. 
 
 Evaluating a process matrix is of particular importance in 
 experimental situations in order to determine how different is
 the actual quantum process with decoherence from the intended 
 operation. It is also quite interesting to perform QPT in case of 
 non-trivial evolutions such as with time dependent Hamiltonians ($H(t)$
 and/or $[H(t),H(t')]\neq0$). QPT also helps to strengthen the understanding
 of various decoherence models. 
 We obtain these process matrices for the case of widely applicable
  counter-intuitive STIRAP and saSTIRAP processes in a qutrit.

\section{QPT of STIRAP and {\small sa}STIRAP}

Stimulated Adiabatic Raman Passage (STIRAP)
ensures the precise population transfer from one quantum state 
to another \cite{Falci2017}. Considering a three-level quantum system with basis 
states $\{|0\rangle, |1\rangle, |2\rangle\}$,
the STIRAP process is exploited here to transfer the population
from ground state ($|0\rangle$) to the second excited state ($|2\rangle$).
In the light of selection rules that forbid a direct $|0\rangle-|2\rangle$
transfer, this process turns out to be quite useful. 
STIRAP realizes an adiabatic evolution of the system initialized in 
state $|0\rangle$ to final state $-|2\rangle$ using two Gaussian
drives $\Omega_{01}$ and $\Omega_{12}$ acting in a counter-intuitive temporal order.
During the evolution, the system at any time $t$ is in instantaneous
 eigenstate ($|D\rangle= \cos\Theta |0\rangle - \sin \Theta |2\rangle $) 
 of the STIRAP Hamiltonian, where $|D\rangle$ is called the dark state
and $\Theta=\tan^{-1}(\Omega_{01}(t)/\Omega_{12}(t))$ is the mixing angle.
The effective STIRAP 
Hamiltonian \cite{kumar-nature-2016} is obtained by applying the rotating-wave approximation (RWA)
\begin{eqnarray} \label{eq:stirap}
\rm{H}_0 &=& \frac{\hbar}{2}  \Omega_{01}(t) \left[ \cos(\phi_{01}) \Lambda_1 + \sin(\phi_{01}) \Lambda_2 \right] 
+  \frac{\hbar}{2} \Omega_{12}(t) \left[ \cos(\phi_{12}) \Lambda_6 + \sin(\phi_{12}) \Lambda_7 \right] \nonumber \\
&-& \frac{\hbar}{2} \left[ \delta_{01} \Lambda_3 + \frac{\delta_{01}+2\delta_{12}}{\sqrt{3}} \Lambda_8 - \frac{2\delta_{01} + \delta_{12}}{3} \mathbb{I}_{3} \right],
\end{eqnarray}
where $0-1$ and $1-2$ drives have Gaussian envelopes with standard deviation $\sigma$ and relative separation $t_s$, given by
\[ \Omega_{01}(t) = \Omega_{01} e^{-t^2/2\sigma^2} \quad \textrm{and} \quad  \Omega_{12}(t) = \Omega_{12} e^{-(t-t_s)^2/2\sigma^2},\] 
with phases $\phi_{01}$ and $\phi_{12}$ respectively. 
These microwave fields may be slightly detuned from the respective transmon transition frequencies $\omega_{01}$ and $\omega_{12}$ by $\delta_{01}$ and $\delta_{12}$ respectively.

Adiabatic processes are demanding in terms of time, which is not always
affordable in real situations. The superadiabatic STIRAP is a protocol
that offers a shortcut to the STIRAP process, in order to have a precise
population transfer in experimentally feasible time scales \cite{Torrontegui13}.
The superadiabatic STIRAP is obtained by adding a counterdiabatic term,
\begin{equation}  
\rm{H}_{cd} =-\frac{\hbar}{2}\Omega_{02}(t) \left[ \cos(\phi_{02}) \Lambda_4 + \sin(\phi_{02}) \Lambda_5 \right],
\end{equation}
which is experimentally realized by a two-photon process~\cite{antti-arxiv-2017}.
Here we obtain an effective Rabi coupling with absolute value $\Omega_{02}(t)=2\dot{\Theta}(t)$ 
and complex phase $\phi_{02}$; also it is convenient to use the notation $\phi_{02} = -\phi_{20}$.
The counteradiabatic condition also implies $\phi_{01}+\phi_{12}+\phi_{20}=-\pi/2$. Based on an analysis of gauge invariance \cite{antti-science-2019,antti-arxiv-2017}, we can take for simplicity $\phi_{01}= \phi_{12} = 0$. Note that for $\phi_{02}=\pi/2$,
the counterdiabatic Hamiltonian $\rm{H}_{cd}$ yields the evolution $\exp\left[i\hbar \Lambda_5\int_{t_i}^t\dot{\Theta}(t) dt\right]$.

This is experimentally 
realized by a two-photon resonance pulse~\cite{antti-arxiv-2017},
\begin{eqnarray} \label{eq:2ph}
\rm{H}_{2\rm{ph}} &=&  \frac{\hbar}{2}\Omega_{2\rm{ph}} \left[ \cos(\phi_{2\rm{ph}}-\Delta t) \Lambda_1 -\sin(\phi_{2\rm{ph}}-\Delta t) \Lambda_2 \right] 
+ \frac{\hbar}{\sqrt{2}} \Omega_{2\rm{ph}} \left[ \cos(\phi_{2\rm{ph}}+\Delta t) \Lambda_6 -\sin(\phi_{2\rm{ph}}+\Delta t) \Lambda_7 \right]. \nonumber \\ \label{eq:h2ph_lambda}
\end{eqnarray}
with the amplitude of the drive $|\Omega_{2\rm{ph}}|=\sqrt{\sqrt{2}\Delta\Omega_{02}}$ and phase $\phi_{2\rm{ph}}=-(\phi_{20}+\pi)/2$,  which simultaneously drives the $|0\rangle-|1\rangle$ and $|1\rangle-|2\rangle$ transitions
with detunings $\mp \Delta$ respectively, where $\Delta=(\omega_1-\omega_2)/2$.
The saSTIRAP Hamiltonian is thus given by
\begin{equation} \label{eq:saSTIRAP}
H_{\rm saSTIRAP}=H_{0}+H_{\rm 2ph}.
\end{equation}

\begin{figure}
	\includegraphics[scale=0.8]{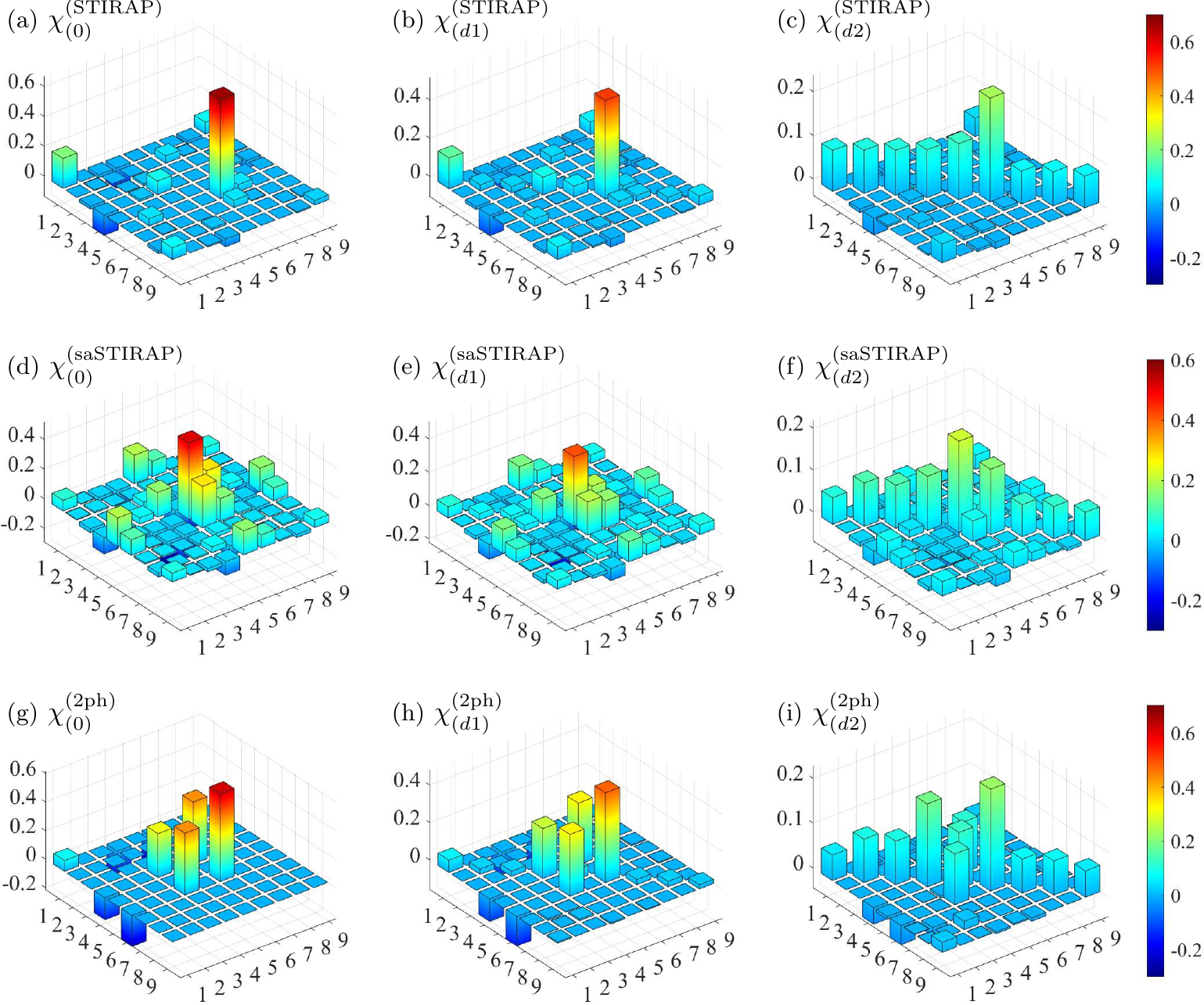}
	% Here is how to import EPS art
	\caption{\label{fig:epsart} The process matrices for (a,b,c) STIRAP, (d,e,f) saSTIRAP,
		(g,h,i) two-photon processes are represented without decoherence, 
		and with two different sets of decoherence parameters 
		labelled by $(d1)$ and $(d2)$.
      Labels $1-9$ along the x and y axes correspond to the set of 
    single-qutrit bases in the operator space, where label $1$ corresponds 
    to the identity operator ($\mathbb{I}$) and labels $2-9$ stand for
    the Gell Mann matrices $\Lambda_1-\Lambda_8$ respectively. The colorbars 
    for sub figures in each row are shown at the right of that row. }
	\end{figure}
We characterize STIRAP, saSTIRAP and two-photon resonance 
processes within the limits
of experimentally feasible parameters via quantum process tomography.
We simulate a qutrit with $|0\rangle-|1\rangle$ and $|1\rangle-|2\rangle$ transition frequencies
 $\omega_{01}/2\pi=5.27$ GHZ and $\omega_{12}/2\pi=4.82$ GHz respectively.
The STIRAP and saSTIRAP drives consist of overlapping Gaussian pulses with maximum 
amplitudes $\bar{\Omega}_{01}/2\pi=\bar{\Omega}_{12}/2\pi=45$ MHz and standard
deviation, $\sigma=35$ ns. The relative separation between these Gaussian pulses is
given by $t_s=-0.8 \times \sigma$. We consider a qutrit initialized in nine different initial states, which are then evolved from 
$t_i=-182$ ns to $t_f=140$ ns with $T=t_f-t_i$ in $1800$ time steps.
 The frequencies of the 
Gaussian pulses are taken to be same as the respective qutrit transition 
frequencies.

We also perform quantum process tomography of STIRAP and saSTIRAP 
in the presence of decoherence. The evolution
of the system is governed by the Lindblad master equation, $\dot{\rho}=-i/\hbar [\rm{H},\rho]+\mathcal{L}[\rho]$, where 
$\mathcal{L}[\rho]= \Gamma_{21}\rho_{22}(|1\rangle\langle 1| - |2\rangle\langle 2|) + \Gamma_{10}\rho_{11}(|0\rangle\langle 0| - 
|1\rangle\langle 1|) - \sum_{j,k\in{0,1,2}, j\neq k}\gamma_{jk}\rho_{jk}|j\rangle\langle k|$ is the Lindblad qutrit superoperator \cite{kumar-nature-2016} 
with decoherence rates $\gamma_{10}= \Gamma_{10}/2 + \Gamma_{10}^{\phi}$, $\gamma_{20} = \Gamma_{21}/2 + \Gamma_{20}^{\phi}$, and $\gamma_{21} = (\Gamma_{10}
+ \Gamma_{21})/2 + \Gamma_{21}^{\phi}$.
We simulate the evolution of this equation with two different 
sets of relaxation rates.
The first set, labelled by $(d1)$, has $\Gamma_{10}=0.5$ MHz and $\Gamma_{21}=0.71$ MHz and pure dephasing rates 
$\Gamma_{10}^{\phi}=0.4$ MHz, $\Gamma_{21}^{\phi}=0.56$ MHz, and $\Gamma_{20}^{\phi}=0.96$ MHz, which correspond to realistic experimental values for transmons. The second set of decoherence parameters, labelled by $(d2)$, simulate
a situation with much higher decoherence. The parameters are
$\Gamma_{10}=2.5$ MHz, $\Gamma_{21}=3.55$ MHz,
$\Gamma_{10}^{\phi}=2$ MHz, $\Gamma_{21}^{\phi}=2.80$ MHz, and $\Gamma_{20}^{\phi}=4.8$ MHz.
The process matrices resulting from these simulations are shown in Figs.~\ref{fig:epsart}(a,b,c) for STIRAP with $H=H_0$ (Eq.~\ref{eq:stirap}), Figs.~\ref{fig:epsart}(d,e,f) 
for saSTIRAP governed by $H=H_{\rm saSTIRAP}$ (Eq.~\ref{eq:saSTIRAP}) and Figs.~\ref{fig:epsart}(g,h,i) for the process
resulting from two-photon resonance drive $H=H_{\rm 2ph}$ (Eq.~\ref{eq:2ph}).
Figs.~\ref{fig:epsart}(a,d,e) do not involve decoherence, 
Figs.~\ref{fig:epsart}(b,e,h) result from the first set of 
typical experimental parameters determining decoherence, and 
Figs.~\ref{fig:epsart}(c,f,i) present the process matrices with 
much higher decoherence which has a higher tedency to enhance the 
elements at the diagonal.

Further, we obtain the fidelity of the process matrices obtained after decoherence
$\chi_{(d)}$ with respect to the one without decoherence $\chi_{(0)}$ by employing the Uhlmann-Josza fidelity
\[ \mathcal{F}_{0,d}= \left[{\rm Tr}\sqrt{\sqrt{\chi_{(0)}} \chi_{(d)} \sqrt{\chi_{(0)}}} \right]^2 .\]
Interestingly, comparing the process matrices for STIRAP and saSTIRAP, the
saSTIRAP process seems to be relatively more robust against decoherence.
This can be observed from the respective fidelities between 
$\chi_{(0)}$, $\chi_{(d1)}$ represented by $\mathcal{F}_{0,d1}$ 
and 
$\chi_{(0)}$, $\chi_{(d2)}$ represented by $\mathcal{F}_{0,d2}$ in 
Table~\ref{table1}.
As another measure for observing  how different the processes are with and without decoherence, we calculate the trace distance between the respective process matrices
by
\[ \mathcal{D}_{\rm 0,d}= \sqrt{{\rm Tr}\left[\chi_{(0)}- \chi_{(d)}\right]^2}. \]
Using this we obtain the trace distances: $\mathcal{D}_{0,d1}$ and 
 $\mathcal{D}_{0,d2}$ between $\chi_{(0)}$, $\chi_{(d1)}$ and
 $\chi_{(0)}$, $\chi_{(d2)}$ respectively.
 
 \begin{table}
 \begin{center}
 \begin{tabular}{|c|c|c|c|c|c|c|c|}
 %	\begin{center}
 	\hline
 &  $\mathcal{F}_{0,d1}$ &  $\mathcal{F}_{0,d2}$ &  $\mathcal{D}_{0,d1}$ &  $\mathcal{D}_{0,d2}$ & $F_0$ & $F_{d1}$ & $F_{d2}$ \\
 \hline 
 STIRAP   & 0.76 & 0.31 & 0.25 & 0.74 & 0.916 & 0.796 & 0.464 \\
 \hline 
 saSTIRAP & 0.78 & 0.33 & 0.24 & 0.72 & 0.999 & 0.861 & 0.487 \\
 \hline 
 Two-photon process & 0.78 & 0.33 & 0.24 & 0.72 & 0.888 & 0.770 & 0.446 \\
 \hline 
%\end{center}
 \end{tabular}
\end{center}
\caption{Fidelity and trace distance values to compare different processes.\label{table1}}
\end{table}
The fidelity $F_{0/d} = \langle 2 |\rho_{0/d} |2 \rangle$ with which the final state $\rho_{0/d}$ without/with decoherence matches with the theoretically expected state $|2\rangle$ 
 is highest in case of saSTIRAP
as shown in last three columns of Table~\ref{table1}, where $F_0$,
 $F_{d1}$, and  $F_{d2}$ represent the quantum state fidelities obtained
 in case of no decoherence, decohering environment with parameter set $d1$
 and decoherence via parameter set $d2$ respectively.
Thus saSTIRAP retains the precision of adiabatic STIRAP as well as the 
robustness of the two-photon drive.

\section{Discussion}
The set of $d \times d$-dimensional operators,
which span the complete operator space of a $d$-dimensional quantum 
system, also serve as a very convenient basis for a process matrix.
Here, in case of $3$-dimensional quantum system, a set of eight Gell-Mann
matrices alongwith the identity operator serve as the basis for the process
matrix and as the generator of various unitary transformations.
In the simplest case of one of the generators $\Lambda_i$ as the
Hamiltonian, there is a direct mapping of $\Lambda_i$ to the $(i+1)^{\rm th}$
diagonal element of the process matrix, while involvement of two or more
 generators give rise to the cross terms in the process matrix. Interestingly,
 the STIRAP Hamiltonian involves terms containing $\Lambda_{1}$, $\Lambda_{2}$
  $\Lambda_{3}$, $\Lambda_{6}$, $\Lambda_{7}$, $\Lambda_{8}$, and $\mathbb{I}$,
  while it is designed to effectively obtain the $0-2$ transition, which in case of direct transition would correspond to $\Lambda_{4}$ and/or  $\Lambda_{5}$. Characterizing the STIRAP process clearly presents the 
  dominated  $\Lambda_{5}$ term in Fig. ~\ref{fig:epsart}(a). Therefore, despite the STIRAP Hamiltonian being non-trivially hiding the $0-2$ 
  coupling, the quantum process tomography correctly identifies the actual 
  process. The saSTIRAP process matrix shown in Fig.~\ref{fig:epsart}(d)
  seem to have additional terms, due to the  addition of the  superadiabatic part of the Hamiltonian.
The process matrices for STIRAP and saSTIRAP with decoherence are shown in Fig.~\ref{fig:epsart}(b,c,e,f) respectively.   We notice that the involvement of decoherence tends to enhance the
   diagonal elements and suppress the cross terms depending upon the 
   decoherence model in simulations or the actual decoherence observed in the
   physical systems.

\begin{acknowledgments}
We acknowledge financial support from Foundational Questions Institute Fund (FQXi) via the Grant No. FQXi-IAF19-06,
 European Commission project QUARTET
 (grant agreement no.
862644, FET Open QUARTET), and from the Academy
of Finland through the RADDESS programme (project
no. 328193) and the “Finnish Center of Excellence in
Quantum Technology QTF” (project 312296).
\end{acknowledgments}

%\nocite{*}
%\bibliography{majorana-stirap}% Produces the bibliography via BibTeX.

%aipnum4-2.bst 2019-01-14 (MD) hand-edited version of apsrev4-1.bst

%%%%%%%%%%%%%%%%%%%%%%%%
\end{document}